\documentclass[aps,prl,twocolumn]{revtex4}
\usepackage{amsfonts}
\usepackage{amsmath}
\usepackage{amssymb}
\usepackage{graphicx}
\usepackage{color}
\def\beq{\begin{equation}}
\def\eeq{\end{equation}}
\def\beqa{\begin{eqnarray}}
\def\eeqa{\end{eqnarray}}

\def\del{\delta}

\def\d{\mathrm{d}}

\def\etal{{\sl et al.},}

\def\nonum{\nonumber \\}

\def\del{\delta}

\def\nonum{ \nonumber \\}

\def\min{\mathrm{min}}

\begin{document}

\title{Superconductor-insulator transition from phase fluctuations in layered superconducting films. }
\author{Y. Dubi$^{1,2}$, Rudro R. Biswas$^{3,4}$ and Alexander V. Balatsky$^{1,3}$}
\affiliation{$^1$ Theoretical Division, Los Alamos National Laboratory, Los Alamos, NM 87545, USA}
\affiliation{$^2$ School of Physics and Astronomy, Tel-Aviv University, Tel-Aviv, Israel }
\affiliation{$^3$ Center for Integrated Nanotechnologies, Los Alamos National Laboratory, Los Alamos, NM 87545 USA}
\affiliation{$^4$ Department of Physics, Harvard University, Cambridge, MA 02138 USA }
%
%

\begin{abstract}
Phase fluctuations in finite thickness layered superconducting films are studied theoretically. The model is shown to reproduce much
of the experimental findings observed in various systems, mainly a superconductor-insulator transition and an inverse dependence of $T_c$ on thickness,$T_c\sim d^{-1}$. The relevance to disordered thin films is discussed.
\end{abstract}
 \maketitle

The superconducting (SC) state and the insulating state are two extremes of the electronic conduction spectrum. It is thus rather surprising that a direct transition between them exists. However, various SC systems such as amorphous and granular films of various compounds exhibit just that a SC to insulator transition (SIT) tuned by an external parameter, typically an external magnetic field of the nominal film thickness \cite{Goldman_review}. The details of the SIT have been a challenge to the community for more than two decades and still many open question remain, especially regarding the thickness-induced transition. The standard picture is that as the film thickness is reduced the amount of disorder increases, eventually causing enough fluctuations that cooper-pairs cannot survive, and thus the SC state is destroyed. The transition is viewed as a quantum phase transition \cite{Gantmakher,Gantmakher2}.

Here we introduce a very simple model which exhibits an apparent SIT and also captures other experimental finding, such as the dependence of $T_c$ on thickness. While not directly related to any specific experimental system, the phenomenological resemblance between the model and various experiments suggests that the mechanism we propose for the SIT should not be ruled out, if only from the principle of Occam's razor.

Consider a system composed of stacks of truly 2D SC films.
Being two dimensional, the action for the SC phase of each film (if they are independent) is  \beq
S=\frac{K}{2} \int \d^2 r ( \nabla\phi )^2 +S_{\textrm{vortices}}\label{layer-action}~~,\eeq
where $\phi$ is the phase of the order parameter and $K$ is the phase stiffness, which is proportional to the superfluid density, and $S_{\textrm{vortices}}$ describes the proliferation of vortices \cite{KT,B}. This action is known to lead to a Kosterlitz-Thouless-Berezinskii (KTB) transition at a temperature $T_{c}\sim \frac{\pi}{2} K$ . We assume that the only role of disorder is to determine the magnitude of the phase stiffness (the stronger the disorder the smaller the stiffness), and we do not assume any electronic inhomogeneity in the system (which may arise with an applied perpendicular field \cite{Dubi}). Since $T_c$ is determined by $K$ we omit the $S_{\textrm{vortices}}$ term from the following actions and consider the effect of thickness on $K$.

The next step is to assume some form of coupling between the phases of the layers. The simplest form which, as we show below, can be solved exactly, is
\beqa
S &=& \sum_i \int \d^2 r \left[ \frac{K_0}{2} (\nabla \phi_i(r))^2+J(\phi_i(r)-\phi_{i-1}(r))^2+\right. \nonum & & \left.-J_1 \nabla \phi_i(r) \cdot \nabla \phi_{i-1}(r) \right] ~,\label{full_action} \eeqa
where $\phi_i(r)$ is the phase of the order parameter in the $i$-th layer, $J$ describes a a discrete generalization of the gradient term to the z-direction, and $J_1$ is a Bio-Savart type of interaction, which (as we show below) emerges naturally from the procedure we use. In an isotropic system one would expect that $J/a^2 \sim K_0$ where $a$ is the distance between layers (in such a way that in a continuum limit this term restores the gradient term in the z-direction), and that $J_1$ be much smaller than $K_0$. Similar actions were previously studied using a renormalization group (RG) approach in the limit of infinite thickness \cite{Benfatto1,Pierson}, where various crossovers were studied as a function of, e.g. the interlayer coupling $J$.

The action of Eq.~(\ref{full_action}) is quadratic, and therefore it is possible to solve for an infinite system. However, we are interested in the effect of finite thickness, i.e. a finite number of layers, which can be obtained as follows. Consider the action of one of the layers, say with an even index $2 i$. Fourier transforming to momentum space, it is given by
\beqa
S_{2i}&=&\sum_{k} \left[ \frac{K}{2} k^2 |\phi_{2i,k}|^2+ \right. \nonum && + J( |\phi_{2i,k}-\phi_{2i-1,k}|^2+|\phi_{2i,k}-\phi_{2i+1,k}|^2)-  \nonum
&& -\frac{1}{2} J_1 k^2( \phi_{2i,k}(\phi_{2i+1,-k}+\phi_{2i-1,-k})+ \nonum && \left. +\phi_{2i,-k}(\phi_{2i+1,k}+\phi_{2i-1,k})) \right]~~,\label{S_2i} \eeqa where $\phi_{i,k}$ is the $k-$th Fourier component of $\phi_{2i}(r)$, and $|\phi_k|^2=\phi_k \phi_{-k}$. It is now straight-forward to integrate out $\phi_{2i,k}$ and obtain an effective action (keeping only second order in $k$) for the odd layers,
\beqa
S &=& \sum_i \int \d^2 r \left[ \frac{\tilde{K}}{2} (\nabla \phi_{2i+1})^2+\tilde{J}(\phi_{2i+1}-\phi_{2i-1})^2-\right. \nonum && \left.-\tilde{J_1} \nabla \phi_{2i+1} \nabla \phi_{2i-1} \right] ~,\label{S_odd} \eeqa
with  $\tilde{K}=K_0+(K_0/2-2 J_1),~\tilde{J_1}=J_1-K_0/4,~\tilde{J}=J/2$. We now repeat the above procedure for the new couplings. This real-space RG procedure yields the following recursion relations,
\beqa
K^{(n+1)}&=& \frac{3K^{(n)}}{2}-2 J_1^{(n)}\nonum
J^{(n+1)}&=&J^{(n)}/2 ~~,
J_1^{(n+1)}=(J_1^{(n)}-\frac{K^{(n)}}{4}) \label{recursion}~~, \eeqa with $J^{(1)}=J,~K^{(1)}=K_0,~J_1^{(1)}=J_1$. The film thickness $d$ is related to the RG index $n$ via $2^n=\frac{d}{a}=\del$.

The recursion equations above can be easily solved. For $J$ we find simply $J^{(n)}=2^{-n}=\left( \del \right)^{-1}$, i.e. it is decoupled from $K$ and $J_1$. The interesting consequence of the recursion relation is in $K^{(n)}$, which is given by $K^{(n)}=\frac{2}{3} 2^{-n} (K_0+4 J_1)+\frac{1}{3} 2^n (K_0-2 J_1)$. For $J_1 \ll K_0$ (but also for the general case), and in the limit of a many layers layer, we find \beq K(d)\simeq \frac{1}{3} K_0 \del \label{K_d} ~~.\eeq

 Before we consider the consequences of Eq.~(\ref{K_d}), Let us first recap the phenomenology observed in typical experiments which exhibit the thickness-induced SIT \cite{Goldman_review,Haviland}. In the experiments, the resistance is measured as a function of temperature for different film thickness. For thick films, thicker than some critical thickness $d_c$, there is a SC transition (i.e. resistance becomes vanishingly small) occurring at some critical temperature $T_c$ which is thickness dependent. For films thinner than $d_c$ the resistance seems to increase with decreasing temperature, never reaching a zero-resistance state. Other than that, most details of this transition (i.e. the resistance at the transition, the existence of a temperature-independent seperatrix, critical exponents etc.), are non-universal and depend on the particular system examined \cite{Haviland,Goldman}. Different Chemical composition, growth and annealing process affect these parameters..

\begin{figure}
\vskip 0.5truecm
\includegraphics[width=8truecm]{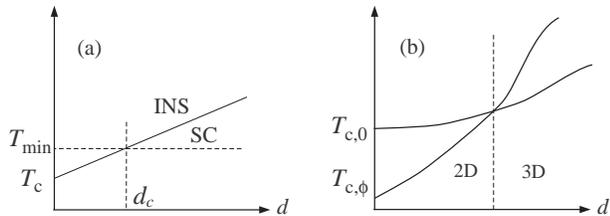}
\caption{(a) The temperature-thickness phase diagram, demonstrating the consequence of a critical temperature which increases with thickness. In the experiments for each measurement the thickness is set and the temperature is lowered down to a finite temperature $T_\min$. Consequently, for films thicker than $d_c$ (Eq.~(\ref{dc})) the temperature will cross $T_c$ and a SIT will take place. For films thinner than $d_c$ the temperature does not reach its critical value and the sample remains in the insulating state. (b)
Dependence of $T_{c,\phi}$ and $T_{c,0}$ on thickness, indicating the crossover from 2D to 3D behavior.}\label{T_d_space}
\end{figure}

 Assume now that the experiments are performed down to a certain minimal temperature $T_\min$ (this is always correct). Since such an experiment does not probe what happens below $T_\min$, Eq.~(\ref{K_d}) means that for films thicker than \beq d_c= \frac{6}{\pi} \frac{a}{K_0} T_\min ~~\label{dc} \eeq the temperatures of the experiment will drop below the film's $T_c$, and it will be considered as a SC film, while thinner films would appear insulating (see Fig.~\ref{T_d_space}(a)) . Above the critical temperature, the resistance of films above and below $d_c$ will be qualitatively similar (see the inset of Fig.~1 of Ref.~\cite{Baturina2}, where the resistances are plotted on a linear scale), and is determined by a complicated interplay of vortices, disorder and finite size effects \cite{Benfatto}. A direct prediction of this argument is that for experiments with increasingly lower temperatures, the resistance of films just below $d_c$ will eventually start decreasing and convert to a SC state. Thus, the critical thickness will increase linearly with $T_\min$ and samples which are currently thought to be insulating will turn out to be superconducting. This implies that the resistance will drop orders of magnitude within a small temperature range in the vicinity of the critical temperature \cite{R-drop}.

  The nature of the true $T=0$ ground state of these films and the possibility of a $T=0$ insulating state due to quantum fluctuations are still open questions. A possible scenario is that quantum fluctuations give rise to an effective minimal temperature $T_Q$ to which the system can reach, which plays the role of $T_\min$ as in the discussion above. This possibility will be explored in the future.

  We point that although we have presented a model which gives the result $T_c\sim d$ analytically, one can also understand this from the experimental data and the fact that in a disordered superconductor the superfluid stiffness is proportional to the mean-free path. The experimental data points that the thickness is simply controlling the mean-free path, implying that the stiffness (and hence the critical temperature) is proportional to the thickness. Thus, the SIT mechanism we propose may be relevant even in the general case which cannot be mapped to a layer model.

From Eq.~(\ref{K_d}) one can also determine the thickness dependence of $T_c$ (see Fig.~2 of Ref.~\cite{Haviland}). Near the transition $K_0$ is approximately given by \cite{Benfatto} $K_0(T) \approx K_0(0) (1-\frac{T}{T_{c,0}})$, where $T_{c,0}$ is the mean-field transition temperature. Inserting this into Eq.~(\ref{K_d}) gives an equation for $T_c$ for a given thickness $d$, $K_0(0) (1-\frac{T_c}{T_{c,0}}) \del=6 T_c/\pi~.$ Solving this yields \beq T_c \approx T_{c,0} \left(1-\frac{6 T_{c,0}}{\pi K_0(0) \del} \right) \label{Tc_d} \eeq for $\del \gg 1$. Such a dependence have been observed long ago \cite{Haviland} and can be extracted from SIT data from various systems, e.g. Bi films \cite{Crauste}, Ge and Xe films \cite{Das Gupta} and even high-Tc films \cite{Matsuda} and seems quite universal. We are, however, unaware of any prior theoretical discussion for the origin of this dependence.


From the above considerations it is simple to understand the presence of a the thermal dimensional crossover \cite{Pierson,Benfatto1}. The system is characterized by two transition temperatures, $T_{c,\phi}$ (quasi-2D KTB transition) and the mean-field transition $T_{c,0}$, both decreasing with decreasing thickness (Fig.~\ref{T_d_space}(b)). As long as $T_{c,\phi}<T_{c,0}$ the system will have a two-dimensional character (with resistance affected by vortices), but for thick enough films the transition will be a regular BCS mean-field type. The transport properties at the crossover region are of interest, and will be a subject of future study.

To summarize, we have introduced a model for layered thin superconducting films. The main result of the model is the observation that the phase-stiffness, which is proportional to the critical temperature, is proportional to the film thickness. A direct consequence of this fact is a possible observation of an apparent SIT, and a dependence of $T_c$ on film thickness given by Eq.~\ref{Tc_d}.

We note that the model may be relevant to other SC systems, such as High-$T_c$ films (where recent progress in fabricating thin films was achieved \cite{Hetel}). These materials are naturally layered, and thus the theory presented here should adequately describe them. However, owing to the high bulk $T_c$, they do not exhibit a T-SIT and remain SC down to very thin samples. We predict that a detailed thickness-dependence study in the extremely under- or over-doped regions should exhibit the same phenomenology as the T-SIT in disordered thin films.

 \acknowledgments
We thank P. Xiong, L. Boulaevskii and J. Zaanen for valuable discussions.  This work was supported by LDRD and  in part, at the Center for Integrated Nanotechnologies, a U.S. Department of Energy, Office of Basic Energy Sciences user facility, by grant No. DE-AC52-06NA25396.

\end{document}